# Modeling Complex Spatial Dynamics of Two-Population Interaction in Urbanization Process

Yanguang Chen[1], Feng Xu[2]

1. Department of Geography, College of Urban and Environmental Sciences, Peking University, Beijing 100871, China. Email: chenyg@pku.edu.cn; 2. China Center for Town Reform and Development, Beijing 100045, China. Email: fengxu.pku@gmail.com

**Abstract:** This paper is mainly devoted to lay an empirical foundation for further research on complex spatial dynamics of two-population interaction. Based on the US population census data, a rural and urban population interaction model is developed. Subsequently a logistic equation on percentage urban is derived from the urbanization model so that spatial interaction can be connected mathematically with logistic growth. The numerical experiment by using the discretized urban-rural population interaction model of urbanization shows a period-doubling bifurcation and chaotic behavior, which is identical in patterns to those from the simple mathematical models of logistic growth in ecology. This suggests that the complicated dynamics of logistic growth may come from some kind of the nonlinear interaction. The results from this study help to understand urbanization, urban-rural population interaction, chaotic dynamics, and spatial complexity of geographical systems.

## 1. Introduction

The study of the logistic equation as viewed from ecology indicates that a simple deterministic system can present periodic oscillation and chaotic behavior along with the model parameter change (May, 1976). However, why the simple model contains complex dynamics still remains ambiguous. Urban study can provide us with facilities for exploring the springhead of complicated dynamics of logistic process. In the urbanization process, the level of urbanization follows the sigmoid curve and can be described with the logistic function (Karmeshu, 1988; United Nations, 1980; United Nations, 1993). Moreover, the urban system and ecological system show comparability in several aspects (Dendrinos, 1992; Dendrinos and Mullally, 1985), which implies that the process of urbanization might have period-doubling bifurcation or chaotic dynamics. In theory, urban system and the process of urbanization can generate complex behaviors such as chaos (e.g. Dendrinos, 1996; Dendrinos and El Naschie, 1994; Nijkamp, 1990; Nijkamp and Reggiani, 1998; Van der Leeuw and McGlade, 1997; Wong and Fotheringham, 1990). Many studies of chaotic cities are relative to spatial interaction and logistic growth.

On the other hand, a great number of simulation analyses and empirical researches show that urban system bears the fractal structure (e.g. Batty and Longley, 1994; Chen and Zhou, 2003; Frankhauser, 1994; White and Engelen, 1994). Fractal structure and chaotic behavior coexist in lots of systems. Fractal property of urban systems suggests complex dynamics of urban evolution. What we concern is not only the bifurcation and chaos in the sheer numerical simulation experiments but also the ones that can be captured from the observation data. One of the viewpoints is that fractal actually appears at the edge of chaos and the coexistence phenomenon of fractal and chaos does not imply the certain correlation between them (Bak, 1996). Perhaps this is true, but we still intend to investigate it from the standpoint of urban systems and urbanization dynamics in order to reveal the relation between the chaotic behavior and fractal structure of nonlinear systems.

Now chaotic cities and fractal cities have become important branch ranges of self-organized cities (Portugali, 2000). *Fractal cities* mean the cities with self-similarity or scaling invariance, while *chaotic cities* suggest the cities with spatial regularity behind random behaviors. The studies of fractal cities and systems of cities are supported by a great number of observations (e.g. Batty and Longley, 1994; Chen and Zhou, 2004; Chen and Zhou, 2006; Frankhauser, 1994; White *et al*, 1997). However, most applications of chaos theory in the social sciences lack empirical content (Nijkamp and Reggiani, 1992). This situation has changed little for more than ten years. In fact, cities and networks of cities are typical complex systems suitable for exploring complicated dynamics (Allen, 1997; Wilson, 2000). The key lies in how to associate theory with practice and reality. The principal aim of this paper is at two aspects. One is to lay an empirical foundation for researching chaotic cities, and the other is to prepare for revealing the essence of complicated behaviors of simple models and the relation between chaotic cities and fractal cities.

The following parts of this paper are structured as follows. In section 2, we build a nonlinear dynamics model about the urban-rural interaction based on the population census data of the United States of America (USA), and then derive the logistic equation of urbanization level from the model. In section 3, with the aid of the US census data, we demonstrate the feasibility and rationality of the model based on statistical analysis, logistic analysis and numerical simulation analysis. This part is used to consolidate the empirical foundation of the model. In section 4, we implement numerical simulation experiment with the model of urbanization dynamics, testifying whether or not such a model presents all the behavior characters of the logistic equation, including periodic oscillation and chaotic behavior. Finally, in Section 5, the discussion is concluded by making some remarks on the significance of the complicated dynamics research from the aspect of the two-population interaction in urban geography.

## 2 Choice and Transform of Mathematical Models

### 2.1 Urban-rural interaction models

A variety of mathematical models has been made to describe the spatial dynamics of the urban-rural population migration. Among these models, two are attention-getting. One is the Keyfitz-Rogers linear model (Keyfitz, 1980; Rogers, 1968), and the other, the United Nations nonlinear model (United Nations, 1980; Karmeshu, 1988). The United Nations adopted a pair of nonlinear equations to characterize the urbanization dynamics

$$\begin{cases} \dfrac{\mathrm{d}r(t)}{\mathrm{d}t} = ar(t) + \varphi u(t) - b\dfrac{r(t)u(t)}{r(t)+u(t)} \\ \dfrac{\mathrm{d}u(t)}{\mathrm{d}t} = cu(t) + \psi r(t) + d\dfrac{r(t)u(t)}{r(t)+u(t)} \end{cases}, \quad (1)$$

where $r(t)$ and $u(t)$ denotes the rural and urban population in time $t$ respectively, $a$, $b$, $c$, $d$, $\varphi$ and $\psi$ are parameters. If parameters $\varphi=\psi=0$, we can derive the logistic model of urbanization level from the UN model. For many years, the United Nations has been using the logistic function to forecast the level of urbanization of each country in the world (United Nations, 1993; United Nations, 2004).

However, empirical studies and statistical analyses show that the urbanization dynamics of many countries such as America, China, and India can be effectively described neither by the Keyfitz-Rogers model nor by the United Nations model. In short, rural population couldn't migrate into urban regions and *vice versa* without spatial interaction between urban and rural population. In other words, population migration and exchange between urban and rural regions depends only on urban-rural population interaction. Consequently, two items of the United Nations model are actually excrescent and equation (1) should be simplified to such a form

$$\begin{cases} \dfrac{\mathrm{d}r(t)}{\mathrm{d}t} = ar(t) - b\dfrac{r(t)u(t)}{r(t)+u(t)} \\ \dfrac{\mathrm{d}u(t)}{\mathrm{d}t} = cu(t) + d\dfrac{r(t)u(t)}{r(t)+u(t)} \end{cases}. \qquad (2)$$

According to equation (2), the rural population can not spontaneously flow into the cities and *vice versa*. The exchange of urban and rural population relies mainly on the urban-rural interaction. The *urban-rural interaction* bears an analogy with the predator-prey interaction in ecology (Dendrinos and Mullally, 1985). The size of urban population is influenced by rural population size and in turn reacts on it. So both the growth rate of urban population and that of rural population depend to a great extent on the coupling or cross correlation between the urban and rural population. For a close region, it is theoretically expected $b=d$. As will be shown later, the US model of urbanization dynamics might be simpler than equation (2). That is $c=0$ in reality.

**2.2 Derivation of the logistic model**

In order to research into the above model, we need to examine it from two ways: one is the logical analysis, and the other empirical analysis. The logical analysis involves at least two aspects. First, whether or not the level of urbanization derived from the above model is close to the logistic increase, and whether or not the total population in a region is limited. Second, whether or not the result of the numerical simulation is coincident with that of the mathematical deduction.

First, we derive the well-known logistic model on the level of urbanization, i.e. *urbanization ratio*. The *level of urbanization* is defined as the proportion or share of urban population in relation to the total population in a region (United Nations, 2004). Thus, we have

$$L(t) = \frac{u(t)}{P(t)} = \frac{u(t)}{r(t)+u(t)} = \frac{V(t)}{1+V(t)}, \qquad (3)$$

where $L(t)$ refers to the level of urbanization, $P(t)=r(t)+u(t)$ to the total population, and $V(t)=u(t)/r(t)$ to the *urban-rural ratio* of population. Differentiating, we get

$$\frac{dL(t)}{dt} = \frac{du(t)/dt}{r(t)+u(t)} - \frac{u(t)}{[r(t)+u(t)]^2}\left[\frac{dr(t)}{dt}+\frac{du(t)}{dt}\right]. \tag{4}$$

Substituting equation (2) into equation (4) yields

$$\frac{dL(t)}{dt} = \frac{cu(t)}{r(t)+u(t)} + \frac{dr(t)u(t)}{[r(t)+u(t)]^2} - \frac{u(t)}{[r(t)+u(t)]^2}\left[ar(t)+cu(t)+(d-b)\frac{r(t)u(t)}{r(t)+u(t)}\right]. \tag{5}$$

For simplicity, taking a region as a close system, then we have $b=d$. In terms of the definition of urbanization level, equation (5) can be transformed into the following form

$$\frac{dL(t)}{dt} = cL(t)[1-L(t)] + (d-a)\frac{r(t)}{u(t)}L(t)^2. \tag{6}$$

According as equation (3), we have an urban-rural ratio

$$V(t) = \frac{u(t)}{r(t)} = \frac{L(t)}{1-L(t)}. \tag{7}$$

This implies $1/V(t)=r(t)/u(t)=1/L(t)-1$. Therefore, equation (6) can be transformed into a logistic equation

$$\frac{dL(t)}{dt} = cL(t)[1-L(t)] + (b-a)\frac{1}{V(t)}L(t)^2 = (b+c-a)L(t)[1-L(t)]. \tag{8}$$

Thus, we have constructed a mathematical relation between models for two interacting population and the logistic growth. Let $k=b+c-a=c+d-a$ represent the *intrinsic rate of growth*. Then equation (8) can be simplified as the usual form

$$\frac{dL(t)}{dt} = kL(t)[1-L(t)]. \tag{9}$$

Solving equation (9) yields the well-known expression of the logistic curve

$$L(t) = \frac{1}{1+(1/L_0-1)e^{-kt}}, \tag{10}$$

where $L_0$ represents the initial value of $L(t)$. That is, when $t=0$, we have $L(t)=L_0$.

A key criterion to judge the urbanization model is the rationality of the increase curve of the total population. Taking derivative of population $P(t)$ with respect to time $t$ gives

$$\frac{dP(t)}{dt} = \frac{dr(t)}{dt} + \frac{du(t)}{dt}. \tag{11}$$

Substituting equation (2) into equation (11) yields

$$\frac{dP(t)}{dt} = ar(t) + cu(t). \quad (12)$$

Obviously, from equation (12) we can get two inconsistent equations as follows

$$\frac{dP(t)}{dt} = cP(t) + (a-c)r(t), \quad (13)$$

$$\frac{dP(t)}{dt} = aP(t) + (c-a)u(t). \quad (14)$$

According to equation (13), when $a>c$, the total population grows more quickly; while according to equation (14), when $a>c$, the total regional population grows slower. These two equations collide with each other. The inconsistency can be eliminated by two conditions: $a=c$ or $c=0$. If $a=c$ as given, then the total population will grow infinitely in the exponential way predicted by Malthus (1798/1996); On the other, if $c=0$, the total population will stop growing when it increases to certain extent. Under the latter circumstance, according as equation (13), since the rural population $r(t) \to 0$, the growth rate of the total population $P(t)$ will gradually decrease to 0; According as equation (14), because the whole population will be completely urbanized, i.e. $u(t) \to P(t)$, the growth rate of the total population will tend toward 0 ultimately. In the real world, we do have $c=0$, as will be illustrated in the following empirical analysis.

It is easy to see that $b$ or $d$ is a very significant parameter in equation (2). On the one hand, it controls the developing trend and quantity of the total population; on the other hand, it affects the original rate of growth $k$ value of the logistic equation on level of urbanization. As we know, parameter $k$ dominates the behavior characters of the dynamical system. When $k>2.57$, the logistic map coming from the discretization of equation (9) will present very complicated behaviors (May, 1976). So what is the case in reality? In the next section, we will validate the above models in virtue of the US observation data. Then we perform numerical simulation experiment to unfold some intrinsic regularity of the urbanization dynamics.

## 3 Empirical Foundation of Two-population Interaction Model

### 3.1 Data and method

The main purpose of this study, as indicated above, is to lay an empirical foundation for further research on complex spatial dynamics of urban-rural interaction. So it is necessary to make relevant statistical analysis of the dynamical equations. There are two central variables in the study of spatial dynamics of urban development: population and wealth (Dendrinos, 1992). According to our theme, we only choose the first variable, population, to test the models. Generally speaking, the population measure falls roughly into four categories: rural population $r(t)$, urban population $u(t)$, total population $P(t)=r(t)+u(t)$, and level of urbanization, $L(t)=u(t)/P(t)$.

The American data comes from the population censuses whose interval is about 10 years. Although the website of American population census offers 22 times of census data from 1790 to 2000, we only use the data from 1790 to 1960 (table 1). The reason is that the US changed the definition of cities in 1950, and the new definition came into effect in 1970. From then on, the American urban population was measured with the new standard. As a result, the statistic caliber of the population data from 1970 to 2000 might be different from those before 1970 although they approximately join with each other (figure 1).

**Table 1 The US rural and urban population and the related data (1790-1960)**

| Time (year) [$t$] | Interval (years) [$\Delta t$] | Rural population [$r(t)$] | Urban population [$u(t)$] | $\dfrac{r(t)u(t)}{r(t)+u(t)}$ | Rural rate of growth [$\Delta r(t)$] | Urban rate of growth [$\Delta u(t)$] |
|---|---|---|---|---|---|---|
| 1790 | 10 | 3727559 | 201655 | 191305.67 | 125855.30 | 12071.60 |
| 1800 | 10 | 4986112 | 322371 | 302794.21 | 172831.00 | 20308.80 |
| 1810 | 10 | 6714422 | 525459 | 487322.03 | 223077.60 | 16779.60 |
| 1820 | 9.8125 | 8945198 | 693255 | 643391.97 | 284153.58 | 44228.48 |
| 1830 | 10 | 11733455 | 1127247 | 1028443.23 | 348484.30 | 71780.80 |
| 1840 | 10 | 15218298 | 1845055 | 1645549.78 | 439908.20 | 172944.10 |
| 1850 | 10 | 19617380 | 3574496 | 3023569.39 | 560942.30 | 264202.20 |
| 1860 | 10 | 25226803 | 6216518 | 4987478.10 | 342920.70 | 368584.30 |
| 1870 | 10 | 28656010 | 9902361 | 7359287.97 | 740346.40 | 422737.40 |
| 1880 | 10 | 36059474 | 14129735 | 10151800.00 | 481402.70 | 797653.00 |
| 1890 | 10 | 40873501 | 22106265 | 14346837.12 | 512383.50 | 810856.70 |
| 1900 | 9.7917 | 45997336 | 30214832 | 18235956.49 | 425582.20 | 1210127.90 |
| 1910 | 9.7917 | 50164495 | 42064001 | 22879255.97 | 163788.26 | 1244862.74 |
| 1920 | 10.25 | 51768255 | 54253282 | 26490822.68 | 221831.22 | 1454372.39 |
| 1930 | 10 | 54042025 | 69160599 | 30336844.29 | 341720.60 | 554473.90 |
| 1940 | 10 | 57459231 | 74705338 | 32478532.68 | 373837.30 | 1542285.60 |
| 1950 | 10 | 61197604 | 90128194 | 36448706.03 | 506197.80 | 2293539.90 |
| 1960 | 10 | 66259582 | 113063593 | 41776788.81 | | |

**Source**: http://www.census.gov/population.

The data displayed in table 1 are fitted to the discretization expressions of the United Nations model and the Lotka-Volterra-type model respectively (r.e. Dendrinos and Mullally, 1985; Lotka, 1956; Volterra, 1931). Since the Keyfitz-Rogers model and the American urbanization model are both special cases of the United Nations model, there is no need to try Keyfitz-Rogers model particularly. The parameters of models are made by the least squares computation, which can make the key parameters, slopes, fall into the most reasonable range.

After estimating the model parameters, we should make tests in two ways. One is the well-known statistical test, and the other is the logical test, which is often ignored in practice. If the model fails to pass the statistical test, it has problems such as incomplete or redundant variables, or inaccurate parameter values; if the model cannot pass the logical test, it has structure problem so that it cannot explain the phenomena at present and predict the developing trend in future. Statistical tests can be made in definite procedure, while the logical test needs to be done with the help of mathematical transformation and numerical simulation experiment.

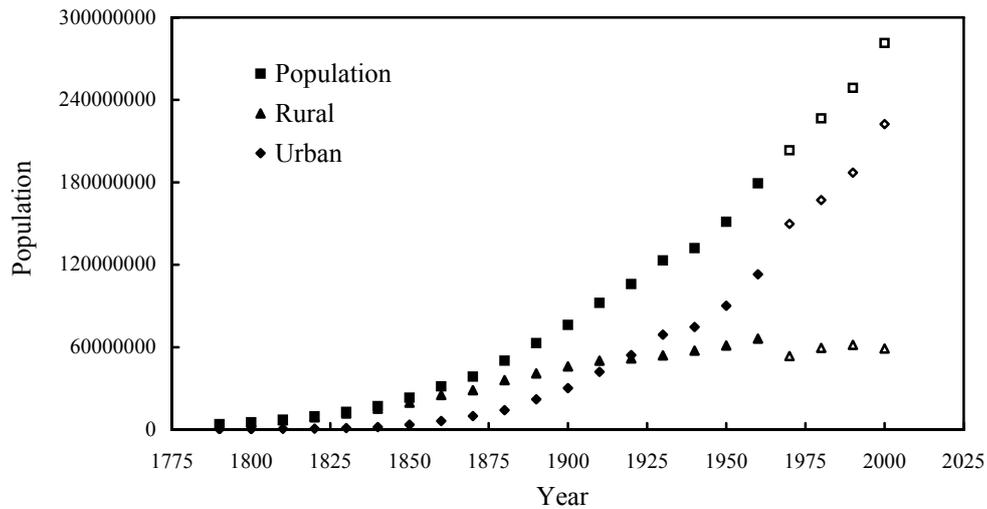

**Figure 1 The changing trend of the US urban, rural and total population (1790-2000)**
(**Notes**: The solid points are data from 1790 to 1960; the hollow points are data from 1970 to 2000. The definition of the city after 1960 is different from before, but the two calibers generally fit with each other.)

## 3.2 Parameters estimation and model selection

In order to make statistical analysis, we must discretize the United Nations model so that it transform from differential equations into difference expressions, i.e., a 2-dimension map. Then the analysis of continuous dynamics changes to that of discrete dynamics. If $\Delta t$=10 as taken, then

d$x$/d$t \propto \Delta x/\Delta t$. Let $r(t)$, $u(t)$ and $r(t)*u(t)/[r(t)+u(t)]$ be independent variables, and $\Delta u(t)/\Delta t$ and $\Delta r(t)/\Delta t$ be dependent variables. A multivariate stepwise regression analysis based on least squares computation gives the following model

$$\begin{cases} \dfrac{\Delta r(t)}{\Delta t} = 0.02584 r(t) - 0.03615 \dfrac{r(t)u(t)}{r(t)+u(t)} \\ \dfrac{\Delta u(t)}{\Delta t} = 0.05044 \dfrac{r(t)u(t)}{r(t)+u(t)} \end{cases}. \qquad (15)$$

This is a pair of difference equations of which all kinds of statistics including $F$ statistic, $P$ value (or $t$ statistic), variance inflation factor (VIF) value and Durbin-Watson (DW) value can pass the tests at the significance level of $\alpha$=0.01 (Appendix 1). In this model, $c$=0. Although we should have $b$=$d$ in theory, they are not equal in the empirical results. There might be two reasons for this. One is that the US is not a truly closed system because of mass foreign migration; the other is that the natural growth of the urban population is dependent on the urban-rural interaction. The second reason might be more important. But on the whole, the equations as a special case of the United Nations model can better describe the American urban and rural population migration process in the recent 200 years.

In light of equation (10), the level of urbanization should follow the logistic curve. It is easy to calculate the urbanization ratio using the data in table 1. A least squares computation involving the percentage urban data gives the following results

$$L(t) = \dfrac{1}{1+20.41573 e^{-0.02238 t}}. \qquad (16)$$

The goodness of fit is $R^2$=0.9839. For convenience, we set $t$=year-1790 (figure 2). Thus we have $k$=0.02238 as the estimated value of the intrinsic growth rate. On the other hand, we could estimate the original rate of growth $k$ value by equation (15): one is $k_1$=$b$-$a$=0.03615-0.02584=0.01031, and the other is $k_2$=$d$-$a$=0.05044-0.02584=0.02460. The intrinsic growth rate should come into between $k_1$=0.01031 and $k_2$=0.03615 and indeed it does. The parameter values estimated from the dynamical system model, equation (15), are similar to that from the logistic model, equation (16). There are some differences between different estimated results due mainly to three factors. The first is non-closed region, the second imprecise data, and the third the computation error resulting from transformation from continuous equation to discrete expression.

For comparison and selection, we also fit the American rural and urban data to the discretization of the predator-prey interaction model. Let *r*(*t*), *u*(*t*) and *r*(*t*)\**u*(*t*) be independent variables and Δ*u*(*t*)/Δ*t* or Δ*r*(*t*)/Δ*t* dependent variables. The multivariable stepwise regression based on least squares computation gives an abnormal result, which cannot be accepted. If we loosen the requirements, then the American urbanization process could be expressed with the Keyfitz model. However, this mathematical expression has two vital shortcomings, which defies us to accept the Keyfitz model for the US urbanization. In short, neither the linear Keyfitz-Rogers model nor the usual non-linear Lotka-Volterra model is as good as the United Nations model in terms of logic sense and statistic effect (Appendix 2).

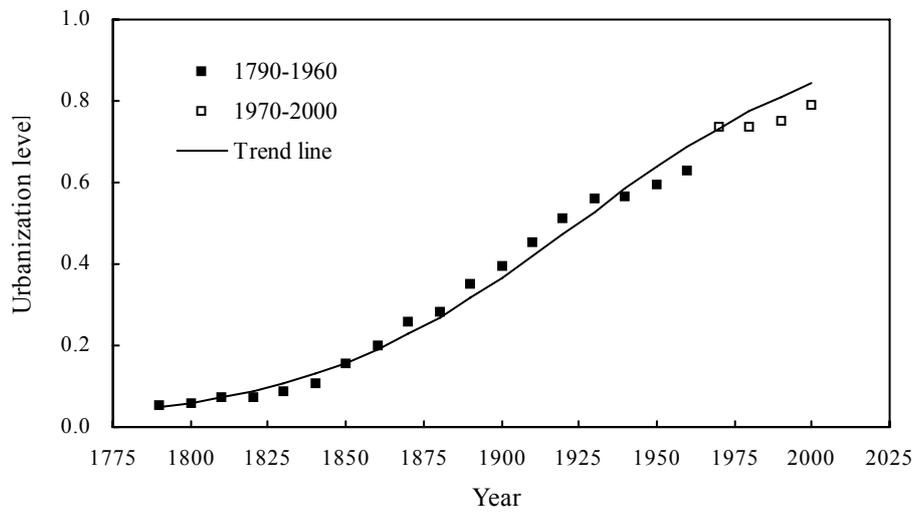

**Figure 2 Logistic process of the US level of urbanization (1790-2000)**
(**Note**: The solid points are the data from 1790 to 1960, and the hollow points the data from 1970 to 2000)

We can generate the data of American urban, rural and total population and the urbanization level by using discrete dynamics model, and then draw a comparison between the simulation value and observed data. Figures 3 and 4 respectively show the simulation results based on equations (15). It is easy to see that the change of the urban and total population approximately follow the path of the S-curve, while the rural population first increases, then decreases, and finally turns itself into the urban population completely (figure 3). Moreover, the level of urbanization increases in the logistic way (figure 4). The changing trend of the numerical simulation results displayed in figures 3 and 4 is roughly coincident with the actual observation data (figures 1 and 2). Although it is unpractical that the saturation value of the urbanization level is 100%, the characters of evolvement of the urban and rural population reflected by the discrete dynamical model, i.e., equation (15), comply with logic rules well. The total population converges, and the change of the percentage urban conforms to the logistic curve.

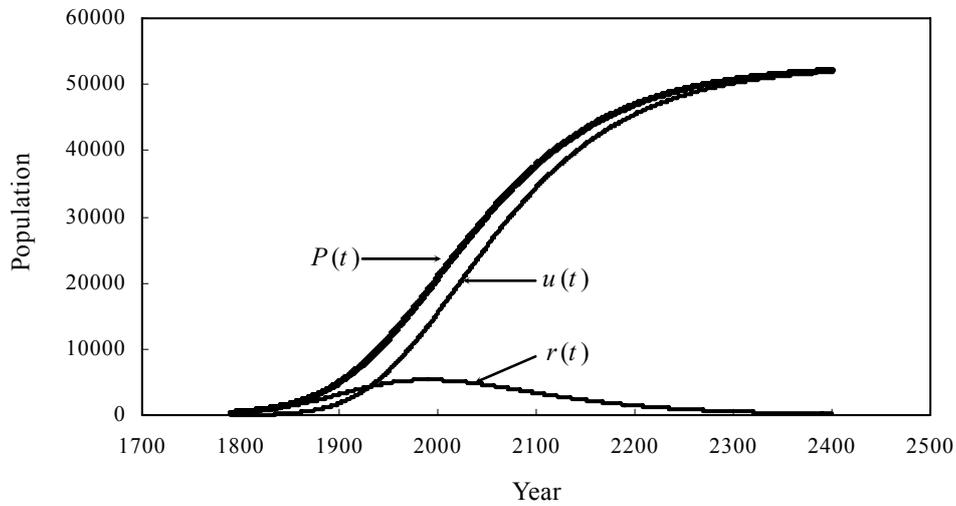

**Figure 3 Numerical simulation curve of rural, urban, and total population in the American urbanization process**

(**Notes**: The numerical simulation results are based on the discrete dynamical equations of urbanization, equation (15), the unit of population is taken as 10,000 persons)

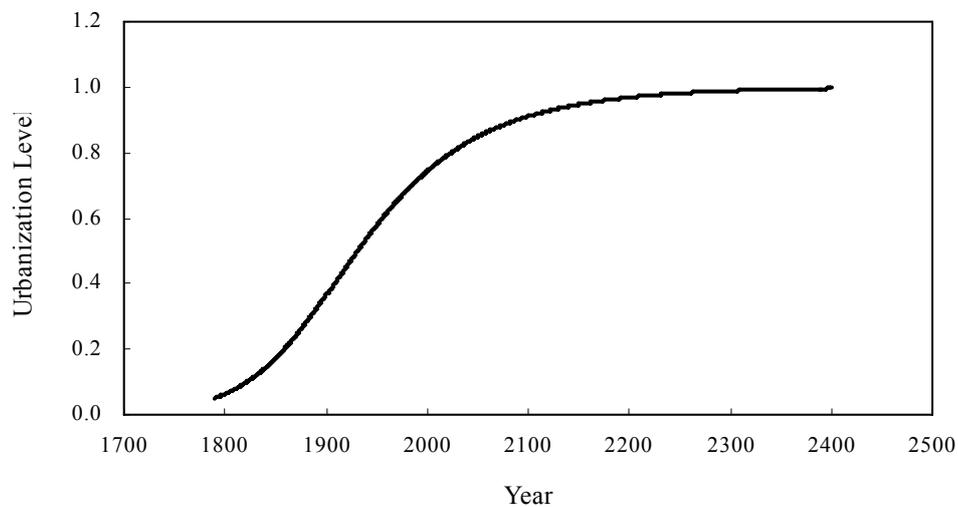

**Figure 4 Numerical simulation curve of American urbanization level (1790-2400)**

(**Notes**: The numerical simulation based on equation (15). The saturation value is 1. The curve is identical in shape to that of logistic growth indicated by equation (16))

To sum up, the American model of urban-rural population interaction can be expressed by equation (2) but the parameter $c=0$. This is the experimental foundation of theoretical analysis of discrete urbanization dynamics. So far, we have finished the building work of the model of urbanization based on the population observation in the real world. In the following section, we will discuss the complicated behaviors of the above model of urbanization dynamics in the possible world in theory.

## 4 Complex Behaviors of Urbanization Dynamics

One of the purposes of this work is to prepare for revealing the essence of complicated behaviors of simple models. The discrete equations of two-population interaction between urban and rural systems can exhibit all the complex dynamics arising from the logistic map, including period-doubling bifurcation and chaos. Chaos theory is a field on the random behavior and latent order of certain dynamical systems, which are highly sensitive to initial conditions (Malanson, *et al*, 1990). Chaos is often defined as intrinsic unpredictability of deterministic systems. In other words, a difference equation is regarded as a chaotic system if the solution to the equation is sensitively dependent on its initial conditions.

The discrete urban-rural interaction model can show richer details of complicated behaviors than what logistic map does, and especially, it can offer a new way of looking at complex dynamics of simple mathematical models. According as equation (15), the parameter $c$=0, thus equation (8) can be reduced to

$$\frac{dL(t)}{dt} = (b-a)L(t)[1-L(t)] = kL(t)[1-L(t)], \qquad (17)$$

where the intrinsic rate of growth is $k=b-a$. The discretization of equation (17) is a finite-difference equation

$$L_{t+1} = (1+k)L_t - kL_t^2, \qquad (18)$$

Defining a new variable $x_t=kL_t/(1+k)$, we can turn equation (18) into the familiar parabola, i.e., a 1-dimension map $x_{t+1}=(1+k)x_t(1-x_t)$.

As we know, according to May (1976), the quadratic map can present periodic oscillation and even more complicated chaotic behaviors under certain conditions. Since equation (17) is derived from equation (2), the behavior characters of equation (18) should be able to be produced by the discretization of equation (2). For testing this hypothesis, we can perform some numerical simulation experiment by using equation (2), which can be discretized as a 2-dimension map

$$\begin{cases} r(t+1) = (1+a)r(t) - b\dfrac{r(t)u(t)}{r(t)+u(t)} \\ u(t+1) = (1+c)u(t) + d\dfrac{r(t)u(t)}{r(t)+u(t)} \end{cases} \quad (19)$$

The conversion between differential equation and difference will result in some subtle change of parameter values. But for simplicity, we don't modify the parameter symbols after converting equation (2) into equation (19). The numerical solutions of equation (19) shows that when the difference between *b* and *a* increases (please notice $k=b-a$), the growth curve of urbanization level $L_t$ indeed changes from simplicity to complexity, from S shape to periodic oscillation and even to chaos. In short, all the behaviors of logistic map revealed by May (1976) can be exhibited by the discrete two-population interaction model (Figure 5).

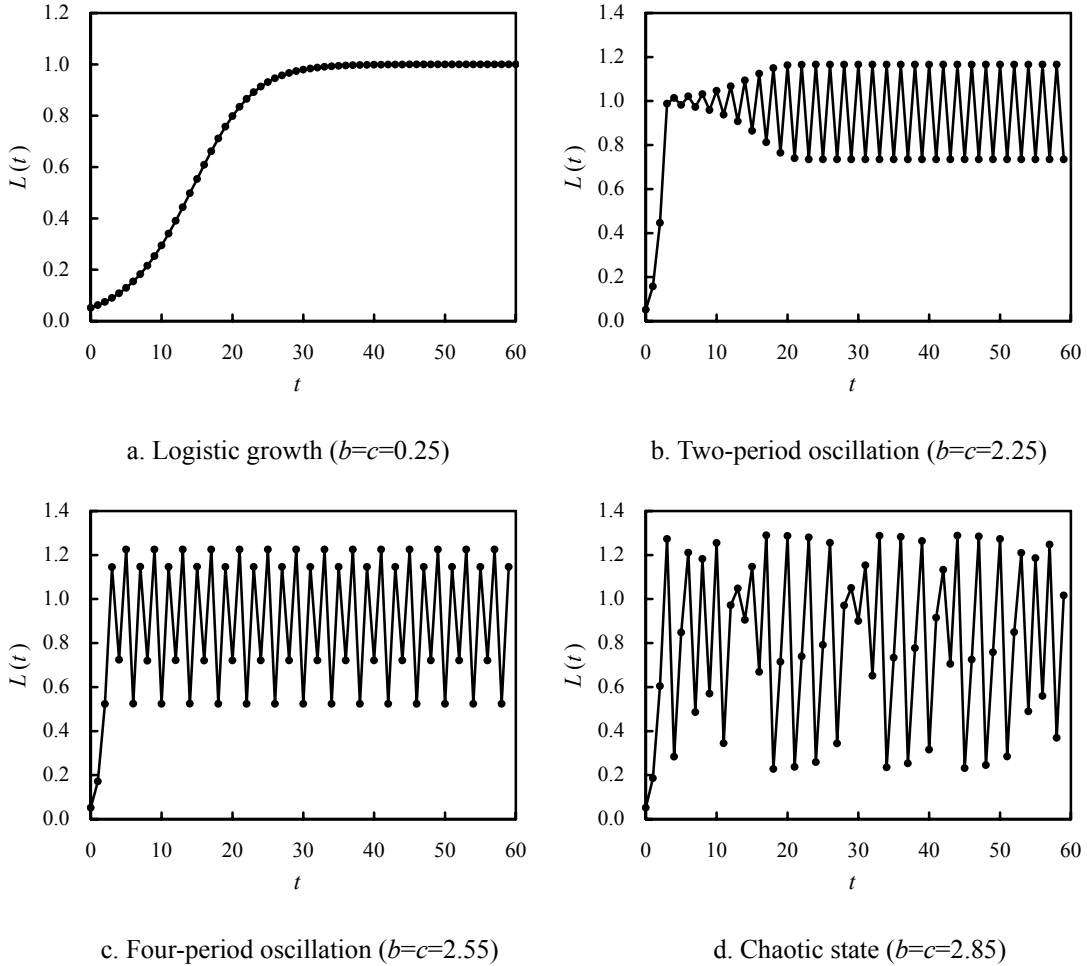

a. Logistic growth ($b=c=0.25$)  
b. Two-period oscillation ($b=c=2.25$)  
c. Four-period oscillation ($b=c=2.55$)  
d. Chaotic state ($b=c=2.85$)

**Figure 5 Four types of changes of urbanization level by urban-rural interaction model: from fixed state to chaos**

(**Notes**: The parameter values are taken as $a=0.025$, $c=0$. In order to correspond to the logistic model, we make $b=d$. The initial urban and rural population values are based on the US census in 1790, i.e., $r(0)=r_0=3,727,559$, $u(0)=u_0=201,655$. It is easy to see that the numerical simulation results from the two-population interaction model are identical in curves to those from the logistic model by May in 1976)

As a matter of conciseness, we may as well set *b*=*d* based on the theoretical hypothesis. According to the estimated results of the US urbanization model, let *a*=0.025 and *c*=0. In addition, the US census data in 1790 are taken as the original urban and rural population values. Then, we increase the value of *b* and *d* continually. The numerical simulation result shows that when *b*=*d*<1.31, the urbanization level presents the S-shaped curve growth, i.e. a fixed state curve; when *b*=*d*>2.025 (*k*=*b*-*a*>2), the dynamical system comes into 2-period oscillation state; when *b*=*d*>2.475 (*k*=*b*-*a*>2.45), the system takes on 4-period oscillation state; then the system will fall into 8, 16, and $2^n$-period state as the values of *b* and *d* increase (*n* is a positive integer); when *b*=*d*>2.6 (*k*=*b*-*a*>2.575), the system will perform random period or chaotic state. The growing limit of parameters is *b*=*d*=3.03. Compared with the work of May (1976), the period-doubling bifurcation route to chaos of the discrete urban-rural interaction model is identical in patterns to that of the logistic map. Of course, there might be subtle difference sometimes. Why the 2-dimension map exhibits the same complex dynamics with that arising from the 1-dimension map? Maybe the two-population interaction model poses a new question about the essence of chaos (Figure 6).

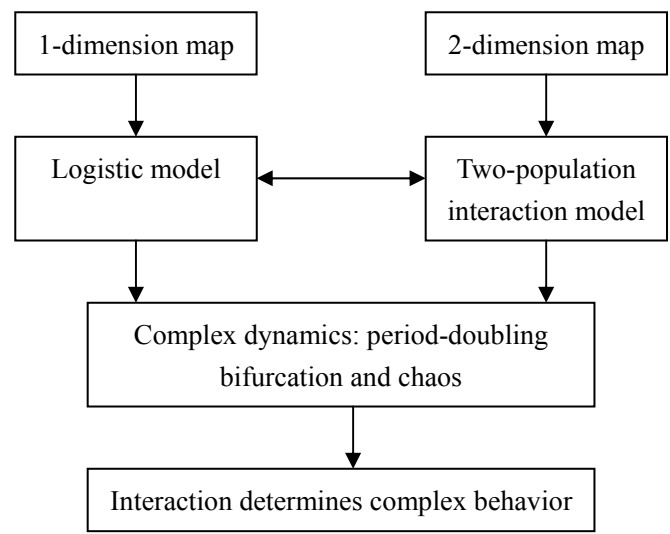

**Figure 6 A 1-dimension map and a 2-dimension map reach the same goal by different routes**

Further, if we ignore the connection between the urban-rural interaction model and the logistic equation by permitting *b*≠*d*, then the behavior features of the dynamical system will become much richer. When we fix *a*, *c*, and *d*, the system will exhibit periodic oscillation or even chaos; however, when we fix *b*, the behavior characters of the system do not change along with the changes of the other parameters. It is obvious that the key parameter that determines the system behavior is *b*, or strictly speaking, is the difference between *a* and *b*. In detail, for instance, let's consider *a*=0.025, *c*=0, and *d*=0.05 according to the aforementioned empirical analysis. The curve of the urbanization level changes along with *b* is in the same way with the result based on *b*=*d*, but the critical values of the period-doubling bifurcation route to chaos increases.

Under the circumstances, when the system comes into the chaotic state, it still presents periodic oscillation. However, the period is not only a multiple of 2 any more, but a random integer. For example, when *b*=3.2, system will enter into period 5 state (figure 7). More experimental results show that system will present period 3 or period 6 in the chaotic state. This illustrates the well-known Sharkovsky's theorem, and remind us of Li and Yorke (1975)'s discovery. The proposition "period three implies chaos" can be expressed equivalently as "period any number beyond 2-multiple implies chaos".

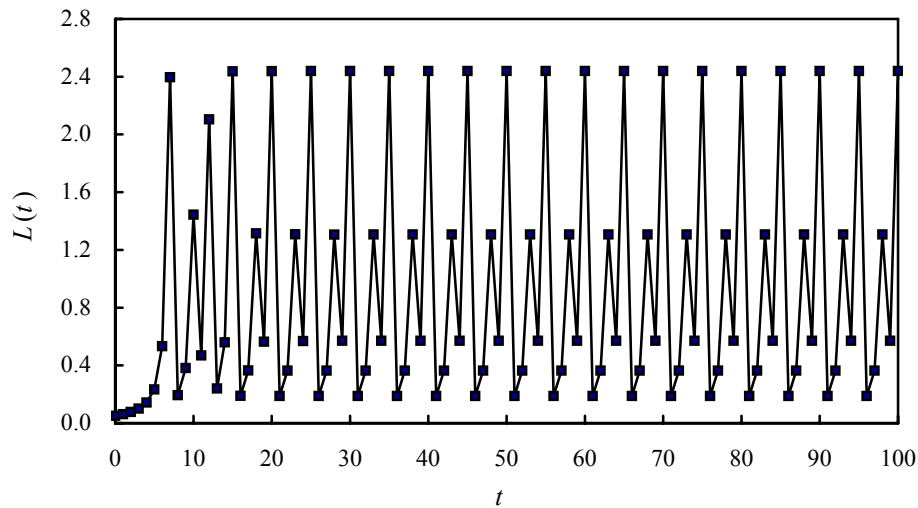

**Figure 7 Period 5 oscillation of urbanization level: a special chaotic state**
(**Note**: The parameter values are such as *a*=0.025, *b*=3.2, *c*=0, *d*=0.05)

Based on the above simulation analysis, we can reach conclusions as follows. First, the key element of the urbanization process lies in rural population rather than urban population. According as the dynamical model of urbanization for America, the non-linear term of urban-rural interaction connects the city on one end and the village on the other. It is the difference of parameters *a* and *b* that dominates the behavior features of the urbanization dynamics. This implies that it is the rural region and urban-rural interaction that determine the progress of urbanization. Secondly, only when *b=d*, there is strict mathematical relation between the urban-rural interaction model and the logistic equation. On the one hand, the logistic model parameter derived by mathematics is *k=b-a*, whose value controls the behavior characters of the logistic map; on the other, numerical simulation shows that it is the value of (*b-a*) that determines the behavior of the urban-rural interaction model. Evidently, the precondition of connecting the urban-rural interaction model with the logistic equation is *b=d*. Third, the periodic oscillation and chaotic behavior of urban-rural interaction maybe only belong to the results of the sheer theoretical analysis. As we know, since the number of the urban and rural population can not be negative in reality, namely $r(t) \geq 0$ and $u(t) \geq 0$, it is certain that $L(t) \leq 1$ in terms of equation (3). However, if the dynamic system exhibits periodic oscillation or even chaotic behavior, the simulation value of the rural population will be negative, namely $r(t)<0$. Thus the numerical simulation shows the abnormal phenomenon of urbanization ratio $L(t)>1$ (see figures 7). Actually, this case is impossible. So period-doubling bifurcation and chaos of urbanization seem not to happen in the real world, it only appears in the imaginary world as a theoretical product (Chen, 2009).

For the sake of understanding the essence of complex dynamics, let's implement a simple mathematical transformation. Removing the nonlinear terms indicating interaction in equation (2) yields

$$\begin{cases} \dfrac{dr(t)}{dt} = ar(t) \\ \dfrac{du(t)}{dt} = cu(t) \end{cases}. \tag{20}$$

The difference between the growth rates of the urban and the rural population is as follows

$$c - a = \frac{du(t)}{u(t)dt} - \frac{dr(t)}{r(t)dt} = \frac{d}{dt}\left[\ln\frac{u(t)}{r(t)}\right]. \tag{21}$$

Taking equation (7) into consideration, we have

$$c - a = \frac{d}{dt}\left[\ln V(t)\right] = \frac{dL(t)}{dt}\left[\frac{1}{L(t)} + \frac{1}{1-L(t)}\right]. \tag{22}$$

Thus equation (22) can be transformed into a logistic equation

$$\frac{dL(t)}{dt} = (c-a)L(t)[1-L(t)]. \tag{23}$$

Although equation (20) also leads to logistic equation, the dynamical patterns of the linear differential equations are very simple. The numerical simulation based on the discretization of equation (20) shows no periodic oscillation, say nothing of chaos. This suggests that complicated dynamics such as period-doubling bifurcation and chaos coming from the logistic map is in fact rooted in interaction associated with nonlinearity.

An allometric scaling relation between urban population and rural population can be derived from equation (20) such as

$$u(t) = \eta r(t)^{b/a} = \eta r(t)^{D_u/D_r} = \eta r(t)^\sigma, \tag{24}$$

in which $\eta = u_0 r_0^{-b/a}$ is a proportionality coefficient, and here $r_0$ and $u_0$ denote the initial values of rural and urban population respectively. Apparently, the allometric scaling exponent $\sigma$ is given by

$$\sigma = \frac{b}{a} = \frac{D_u}{D_r}, \tag{25}$$

where $D_u$ refers to the fractal dimension of urban population, and $D_r$ to the fractal dimension of rural population. This suggests that the logistic equation and the related transformation may be the mathematical link between the chaos and fractals of urban evolvement.

Another interesting discovery is that the logistic equation comes between the exponential growth models indicating simplicity, equation (20), and the two-population interaction model indicating complexity, equation (2). This reminds us that the logistic model maybe implies a mathematical transform between simple expressions and complex dynamics. As space is limited, it is impossible to make all these questions clear here and the pending questions will be discussed in future reports.

## 5 Conclusions

Urbanization is a complex process of spatial dynamics with two-population interaction. The mathematical model based on the US observation data can be proposed to describe the nonlinear evolvement. From the urban-rural interaction model, we can strictly derive the logistic equation about level of urbanization. As logistic growth exists widely in the nature and human society, the two-population interaction model may reflect a kind of ubiquitous dynamical systems. Therefore, some research conclusions can be generalized to other fields, including ecology, economics, and geology. Furthermore, since the logistic model is related to chaos, this implies that the process of urbanization has the potential possibility to bear periodical oscillation and chaotic behavior. Accordingly, the urban-rural nonlinear interaction models could help us better understand the urban system by chaos theory, and meanwhile comprehend more the nature of chaos by urban evolution. The main conclusions of this article are summarized as follows.

First, the urban-rural population interaction model can exhibit periodical oscillation and chaotic behavior in theory. The period-doubling bifurcation diagram route to chaos is identical in patterns to that from the logistic model in ecology. This provides us a new perspective to understand the complicated dynamics of simple systems. The numerical simulations show that the urban-rural population interaction model can exhibit 2-period oscillation, 4-period oscillation, …, $2^n$-period oscillation, and finally chaos, along with the change of parameter values. The changing regularity reminds us of the logistic map. Further research suggests that the discrete two-population interaction model can tell us more about the complex systems than what the logistic model does.

Secondly, complicated dynamics such as period-doubling bifurcation and chaos result from interaction instead of sheer logistic processes. The logistic model on the level of urbanization can be derived not only from urban-rural population interaction models, but also from the allometric equations of rural and urban population growth. However, the behavior patterns of the exponential growth models are very simple. In other words, periodical oscillation and chaotic behaviors can never be generated by means of the exponential growth equations. This implies that no complicated dynamics appears without interaction between rural and urban population.

Thirdly, the logistic equation may possibly form a mathematics transform relation between simplicity and complexity, which provide us a new way to look at complexity. As indicated above, the logistic model can be derived from the urban-rural interaction models or from a pair of exponential equations. This suggests that the logistic equation may act as the mathematical transform from the nonlinear interaction models to the simple linear equations. Such a relation may become a bridge connecting simplicity and complexity. In particular, if such transform can be testified to have universal property, it can be developed into a fire-new logistic transform method. If so, the transform will probably associate the unanalysable nonlinear equation with simple rules and take the opportunity to solve some long-standing unanalysable nonlinear problems.

**Acknowledgements:** This research was sponsored by the National Natural Science Foundation of China (Grant No. 40771061).

# Appendices

## A.1 Regression analysis results of the US urbanization model

The regression analysis results of the US model of urbanization based on the least squares computation are tabulated as follows (Table A1, Table A2). The contents include ANVOA summary, estimated values of model's coefficients and related statistics.

Table A1 ANVOA summary of the US urbanization model (case 1)

| Dependent variable | Model | Independent variable | $R^2$ | Std. Error of the Estimate | Durbin-Watson | $F$ | P-value (Sig.) |
|---|---|---|---|---|---|---|---|
| $\Delta r(t)$ | 1 | $r(t)$ | 0.682 | 233020.670 | 0.527 | 34.383 | 2.399E-05 |
|  | 2 | $r(t)$, $r(t)u(t)/P(t)$ | 0.873 | 152286.817 | 1.472 | 51.482 | 1.917E-07 |
|  | 3 | $r(t)$, $r(t)u(t)/P(t)$, $u(t)$ | 0.946 | 102505.113 | 3.116 | 82.122 | 3.986E-09 |
| $\Delta u(t)$ | 1 | $r(t)u(t)/P(t)$ | 0.905 | 295964.057 | 1.756 | 152.865 | 1.333E-09 |

Table A2 Coefficients and related statistics of the US urbanization model (case 1)

| Dependent variable | Model | Independent variable | Regression Coefficients | Std. Error | $t$ | P-value (Sig.) | VIF |
|---|---|---|---|---|---|---|---|
| $\Delta r(t)$ | 1 | $r(t)$ | 0.00908 | 0.00155 | 5.864 | 2.399E-05 | 1 |
|  | 2 | $r(t)$ | 0.02584 | 0.00368 | 7.026 | 4.100E-06 | 13.201 |
|  |  | $r(t)u(t)/P(t)$ | -0.03615 | 0.00763 | -4.739 | 2.635E-04 | 13.201 |
|  | 3 | $r(t)$ | 0.04128 | 0.00431 | 9.572 | 1.602E-07 | 40.048 |
|  |  | $r(t)u(t)/P(t)$ | -0.14059 | 0.02444 | -5.753 | 5.000E-05 | 299.145 |
|  |  | $u(t)$ | 0.03419 | 0.00782 | 4.371 | 6.393E-04 | 145.326 |
| $\Delta u(t)$ | 1 | $r(t)u(t)/P(t)$ | 0.05044 | 0.00408 | 12.364 | 1.333E-09 | 1 |

When we take the rate of growth of the rural population as dependent variable, a least squares computation yields the following alternative model

$$\frac{\Delta r(t)}{\Delta t} = 0.04128 r(t) - 0.14059 \frac{r(t)u(t)}{r(t)+u(t)} + 0.03419 u(t).$$

If this model is employed to describe the growth of the rural population, the saturation value of the urbanization ratio will be less than 1, which tallies with the actual situation better. However, this model gives rise to two problems. First, the model cannot avoid multi-collinearity, which could be detected from the VIF value in Table A2. Second, based on the model, the total population will not converge but increase infinitely.

**A.2 Regression analysis results based on the Lotka-Volterra model**

The multivariable stepwise regression based on least squares computation gives the following expressions

$$\begin{cases} \dfrac{\Delta r(t)}{\Delta t} = 0.03166 r(t) - 0.07763 u(t) + 0.0000000010113 r(t)u(t) \\ \dfrac{\Delta u(t)}{\Delta t} = 0.01567 r(t) + 0.00000000016266 r(t)u(t) \end{cases}.$$

The first equation has serious problems. Firstly, the estimated values of the parameters cannot pass logical test. The coefficient of the linear term, $u(t)$ should be positive, but it is here negative. The physical meaning of the negative coefficient is inexplainable. What is more, the coefficient of the non-linear term, i.e., the cross term, $r(t)u(t)$, should be negative, indicating that the urban-rural interaction can transform the rural population into the urban population, but it is positive here. This conflicts with the symbol of the non-linear term of the second equation, in which the coefficient value of the cross term is positive, too.

Secondly, some values fail to pass the statistic test either. VIF value is far more than 10, which implies that there exists serious multi-collinearity between the three independent variables. If we eliminate the non-linear term, then new problems will rise. The test of serial correlation of residual errors can not be acceptable ($DW$=1.082), and the symbol problem of the linear term $u(t)$ remains unresolved. If we further remove the linear term $u(t)$, then the $DW$ value will decrease to 0.527, which is more unacceptable (Table A3, Table A4). This suggests that the variables are insufficient, or the serial correlation is serious, either of which is against the basic rules of regression modeling. As for the second equation, nothing seems wrong statistically, but it cannot be understood in logic. According to this equation, the rural population will automatically flow into cities without the urban-rural interaction and the urban population will increase without any relation to itself.

**Table A3 ANVOA summary of the US urbanization model (case 2)**

| Dependent variable | Model | Independent variable | $R^2$ | Std. Error of the Estimate | Durbin-Watson | $F$ | P-value (Sig.) |
|---|---|---|---|---|---|---|---|
| $\Delta r(t)$ | 1 | $r(t)$ | 0.682 | 233020.670 | 0.527 | 34.383 | 2.399E-05 |

| Dependent variable | Model | Independent variable | | | | | |
|---|---|---|---|---|---|---|---|
| | 2 | r(t), u(t) | 0.819 | 181628.502 | 1.082 | 33.965 | 2.694E-06 |
| | 3 | r(t), u(t), r(t)u(t) | 0.889 | 147333.597 | 2.225 | 37.343 | 6.232E-07 |
| $\Delta u(t)$ | 1 | r(t) | 0.887 | 322617.571 | 1.402 | 126.115 | 5.344E-09 |
| | 2 | r(t), r(t)u(t) | 0.919 | 282511.849 | 1.913 | 85.165 | 6.473E-09 |
| | 3 | r(t), u(t) | 0.914 | 290705.035 | 1.897 | 80.015 | 9.939E-09 |

**Table A4 Coefficients and related statistics of the US urbanization model (case 2)**

| Dependent variable | Model | Independent variable | Regression Coefficients | Std. Error | t | P-value (Sig.) | VIF |
|---|---|---|---|---|---|---|---|
| $\Delta r(t)$ | 1 | r(t) | 0.00908 | 0.00155 | 5.864 | 2.399E-05 | 1 |
| | 2 | r(t) | 0.01854 | 0.00306 | 7.026 | 2.168E-05 | 6.413 |
| | | u(t) | -0.00980 | 0.00291 | -4.739 | 4.237E-03 | 6.413 |
| | 3 | r(t) | 0.03166 | 0.00507 | 9.572 | 2.152E-05 | 26.806 |
| | | u(t) | -0.07763 | 0.02299 | -5.753 | 4.520E-03 | 607.776 |
| | | r(t)u(t) | 1.011E-09 | 3.410E-10 | 4.371 | 1.022E-02 | 419.272 |
| $\Delta u(t)$ | 1 | r(t) | 0.02409 | 0.00214 | 11.230 | 5.344E-09 | 1 |
| | 2 | r(t) | 0.01567 | 0.00395 | 3.967 | 1.241E-03 | 4.424 |
| | | r(t)u(t) | 1.62657E-10 | 6.716E-11 | 2.422 | 2.858E-02 | 4.424 |
| | 3 | r(t) | 0.01433 | 0.00489 | 2.928 | 1.038E-02 | 6.413 |
| | | u(t) | 0.01011 | 0.00466 | 2.169 | 4.654E-02 | 6.413 |

If we loosen the requirements, then the American urbanization process could be expressed with the Keyfitz model such as

$$\begin{cases} \dfrac{\Delta r(t)}{\Delta t} = 0.00908 r(t) \\ \dfrac{\Delta u(t)}{\Delta t} = 0.01433 r(t) + 0.01011 u(t) \end{cases}.$$

This mathematical expression has two vital shortcomings. The first is the logical problem. According as the model, urban population, rural population and the total population will increase exponentially without any limit, which is against our common sense. The second is the statistic problem. That is, the second equation can not pass the DW-test (Table A3, Table A4).